\documentstyle[12pt,epsf]{article} 
\def\to{\rightarrow}
\def\ee{e^+e^-}
\def\WW{W^+W^-}

\def\lapproxeq{\lower .7ex\hbox{$\;\stackrel{\textstyle
<}{\sim}\;$}}
\def\gapproxeq{\lower .7ex\hbox{$\;\stackrel{\textstyle
>}{\sim}\;$}}

\makeatletter
\def\slash{\@ifnextchar[{\fmsl@sh}{\fmsl@sh[0mu]}}
\def\fmsl@sh[#1]#2{%
  \mathchoice
    {\@fmsl@sh\displaystyle{#1}{#2}}%
    {\@fmsl@sh\textstyle{#1}{#2}}%
    {\@fmsl@sh\scriptstyle{#1}{#2}}%
    {\@fmsl@sh\scriptscriptstyle{#1}{#2}}}
\def\@fmsl@sh#1#2#3{\m@th\ooalign{$\hfil#1\mkern#2/\hfil$\crcr$#1#3$}}
\makeatother

\begin{document}
\begin{titlepage}
\begin{flushright}
DTP/96/88\\
hep-ph/9610284\\
October 1996 \\
\end{flushright}
\begin{centering}
\vspace*{1.cm}

{\Large{\bf Supersymmetric QCD corrections to the \\[3mm]
		    $W$-boson width}}\\
\vspace{1.3cm}

{\bf A.\ B.\ Lahanas} $^{a}$ \, and \, {\bf V.\ C.\ Spanos} $^{b}$  \\
\vspace{.8cm}
$^{a}$ {\it University of Athens, Physics Department, 
Nuclear and Particle Physics Section,\\ 
GR--15771  Athens, Greece}\\

\vspace{.5cm}
$^{b}$ {\it University of Durham, Department of Physics, \\
Durham DH1 3LE, England}  \\
\end{centering}
\vspace{2.cm}
\begin{abstract}
We calculate the one-loop supersymmetric QCD corrections to the width of 
the $W$-boson. We find that these are of order 
$\sim {{ \alpha_s} \over {\pi}}
{1 \over 20} { {M^2_W} \over {M_S^2}} \Gamma_{u {\bar d}}$, where $M_S$ 
is the
supersymmetry breaking scale and  $ \Gamma_{u {\bar d}}$ the tree level 
hadronic width for   $W^+ \rightarrow u {\bar d} $. Due to the 
appearance of the suppression factor $\sim {1 \over 20}$ these are at least
two orders of magnitude smaller than the standard QCD corrections
$\sim {{ \alpha_s} \over {\pi}} \Gamma_{u {\bar d}}$ 
and hence of the order of the two-loop electroweak effects.
Therefore supersymmetric QCD corrections will only be of relevance once 
experiments reach that level of accuracy.
\end{abstract}

\vspace{3.6cm}
\noindent
\rule[0.cm]{11.6cm}{.009cm} \\     
\vspace{.3cm} 
E-mail: \, $^a$ alahanas@atlas.uoa.gr, $^b$ V.C.Spanos@durham.ac.uk
\end{titlepage}
\newpage
\baselineskip=18pt
The second phase of the LEP (LEP2) collider has already started, and the first
$\ee\to \WW$ events have been collected. Studying for the very first
time directly this process, one will have the opportunity to test the
non-abelian character of the Standard Model (SM), through the precise
measurements of the trilinear gauge boson couplings. In addition, it
will be possible to measure precisely the mass and width of the $W$-boson
\cite{cern}.
Specifically the measurement of the $W$-width is of special interest, as 
it is used as an input parameter in many other processes. (It is understood 
that all the $W$ production events are detected through the hadronic and/or
(semi)leptonic decays of the $W$-boson.) So it is very essential, both
for theoretical and experimental reasons, to know as precise as possible the
theoretical prediction for this parameter. 

The one-loop corrections to the $W$-width in the context of the  
Standard Model (SM) are
already known \cite{chang,denner}, and there has been also a 
calculation in the context of a two Higgs doublet model \cite{shin}.

The possible existence of new physics of characteristic scale $M_{new}$
may affect the theoretical predictions for the $W$-boson decay width.
The magnitude of these effects is not a priori known without knowledge 
of the underlying theory{\footnote{It is known however that there are 
no oblique corrections from new physics in 
$\Gamma(W \to e \nu)$, as pointed out in Ref. \cite{rosner}.}}
and thus manifestation of new physics from a direct measurement of the  
$W$-boson width is not possible. The corrections to the $W$-boson
observables which are induced by new physics are expected to be small, 
possibly smaller than the experimental precision of LEP2 which will
be in the percent region. With increasing experimental accuracy
in the future, the $W$-boson observables may provide a laboratory
for testing new physics and Supersymmetry is a prominent candidate.

It is known that strong interaction effects yield the largest contribution, 
${\cal O}(4\%)$, to the $W$-width at the one-loop order. With the SM being  
promoted to a supersymmetric theory,  the QCD sector is also 
supersymmetrized (SQCD) 
and new species which interact strongly affect the QCD predictions.
Therefore it seems natural to calculate the SQCD corrections to the
hadronic width of the $W$. The size of these corrections depends on the
supersymmetry breaking scale $M_S$ and is obviously negligible as $M_S$
becomes large. However the existing experimental lower bounds on 
sparticle masses does not exclude values of $M_S$ in the vicinity of the 
electroweak scale $M_S \simeq {\cal O}$(few\,$M_W$), in which case these 
effects may not be suppressed.

In this Letter we undertake this problem and calculate the supersymmetric 
QCD corrections to the $W$-boson  width. 
We perform our calculations using the on-shell renormalization 
scheme \cite{on-shell,passa} which has been extensively used in the SM 
calculations (see for instance Ref. \cite{hollik,hollik2}). In order to study
the SQCD corrections to the $W$-boson hadronic width
we need to calculate the corrections to $Wu \bar d$ vertex as well as
the wave function renormalizations to the external fermion propagators
(see Fig. 1(a) and (b) respectively). 
In order to simplify our discussion we shall neglect mixings of the up 
${\tilde u}_L,{\tilde u}_L^c$ and down ${\tilde d}_L,{\tilde d}_L^c$  
left handed squarks of the first two generations since these 
mixings are proportional to the corresponding fermion masses and hence
small. Therefore the above squark states are mass eigenstates in this
approximation.

\begin{figure}[t]
\setlength{\unitlength}{1cm}
\begin{center}
\epsfxsize=19.cm
\epsffile[0 380 670 600]{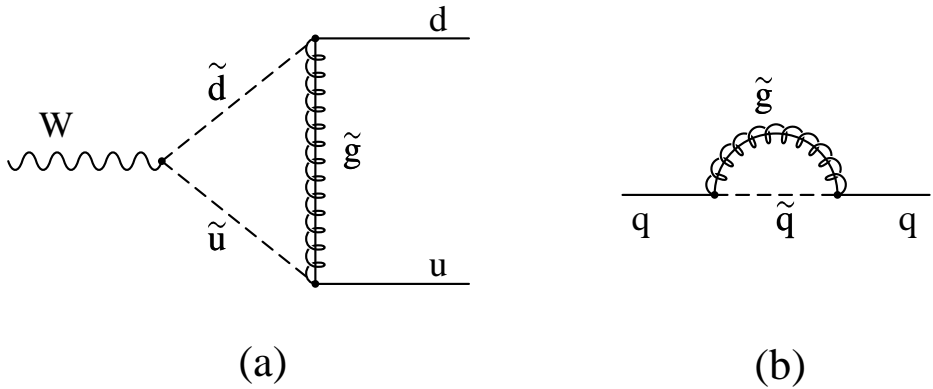}
\begin{minipage}[t]{14.7cm}
\caption[] {Graphs which contribute to the one-loop
supersymmetric QCD corrections to the $W$-width. There are corrections
 to the $Wu\bar d$ vertex (a)
and corrections to the external quark propagators (b).} 
\end{minipage}
\end{center}
\end{figure}
By using the well known Passarino--Veltman functions \cite{passa,hooft}
$B_0, B_1, B_1', C_0, C_1, C_{ij}$ etc., through which the two and three 
point functions are usually expressed{\footnote{In this article we 
follow the convention of Ref. \cite{hollik2} for the definition of 
the Passarino--Veltman functions.}},
we find for the $Wu \bar d$ vertex correction  
of Fig. 1(a)  
\begin{eqnarray}
{\cal A}_1 \,=\,i ({{\alpha_s} \over {2 \pi}}) {g \over \sqrt{2}} \, c_F
\, [ (2 C_{24})\, M_0 + 2 M_u (C_{21}+C_{11}-C_{12}-C_{23}) \,M_2
\nonumber \\
\hspace{3.5cm} + \, 2 M_d (C_{23}+C_{12}) \, M_3 \,]. 
\end{eqnarray}
In the equation above the four basic amplitudes $M_{0,1,2,3}$ are as 
in Ref. \cite{denner}, $g$ is the weak coupling constant 
and $\alpha_s={g^2_s \over (4 \pi)}$, 
where $g_s$ is the strong coupling constant. 
$M_{u,d}$ are the masses of the $u, d$ external quarks and
the factor $c_F = 4/3 $ is the value of 
the quadratic Casimir operator of the fundamental representation
of the SU(3) symmetry group. 
The arguments of the $C_{ij}$ functions appearing above are defined as follows:
\begin{eqnarray}
C_{ij} \,=\, {C_{ij}}(p_1,-p_1-p_2,M^2_{\tilde g},m^2_{{\tilde u}_L},
m^2_{{\tilde d}_L}) .  \nonumber
\end{eqnarray}
In this expression $p_1(-p_2)$ is the momentum carried by the outgoing 
(incoming)
$u(d)$ quark. The ultraviolet infinity of the vertex correction is 
contained within 
the factor $C_{24}$ of the amplitude $M_0$. This infinity is canceled
by the vertex counterterm in the Lagrangian \cite{hollik2}, 
\begin{eqnarray}
{\Delta L}_{CT} \, = \,(\delta Z_L + \delta Z_1^W - \delta Z_2^W ) \,
{g \over \sqrt2} W_{\mu}^{+} {\bar u}_L {\gamma^\mu} {d_L}, 
\end{eqnarray}   
$\delta Z_L \equiv Z_L - 1$, where $Z_L$ is the wave function 
renormalization
constant of the left handed doublet $(u_L, d_L)$. There are no 
strong interaction contributions to the difference
 $  \delta Z_1^W - \delta Z_2^W $ so that only   $\delta Z_L$ needs be 
considered. For the down quark  
the on-shell renormalization condition is
\begin{eqnarray}
{S_{down}}(P)\,\, {\stackrel {{\slash{P}} \rightarrow  M_d }{\longrightarrow}} 
\, ({\slash P } - M_d)^{-1}, \label{yy}  
\end{eqnarray}
where   ${S_{down}}(P) $ denotes the down quark propagator.
This fixes the wave function renormalization constant of both left and
right handed components of the down quark. By a straightforward calculation 
of the graph shown in Fig. 1(b), and using Eq. (\ref{yy}), 
we find for  $\delta Z_L$, which is needed for our calculation,
\begin{eqnarray} 
  \delta Z_L =  ({{\alpha_s} \over {2 \pi}})  \, c_F \,
  [{B_1}(M^2_d, M^2_{\tilde g},m^2_{{\tilde d}_L}) +
   M^2_d \, ( {{B_1}'}(M^2_d, M^2_{\tilde g},m^2_{{\tilde d}_L})
    +  {{B_1}'}(M^2_d, M^2_{\tilde g},m^2_{{\tilde d}_L^c}) )]
\nonumber \\  \equiv   {{\Pi}_d}  (M^2_d).   \label{aa}
\end{eqnarray} 
Therefore the SQCD contribution of the vertex counterterm to 
the  $W^+ \to u \bar d$ amplitude is 
\begin{eqnarray}
{{\cal A}_1}' \,=\, i {g \over \sqrt{2}} M_0 \, ( \delta Z_L ),
\end{eqnarray}  
with  $\delta Z_L $ as given above.

Having fixed the the  renormalization constant $Z_L$ 
it is convenient to  choose the renormalization
constant of the right handed up quark is such a way that the residues
for the left and right handed propagators are equal 
(see for instance Ref. \cite{denner,hollik,hollik2}).
Thus we have for the up quark propagator
\begin{eqnarray}
{S_{up}}(P)\,\, {\stackrel {{\slash{P}} \rightarrow  M_u }{\longrightarrow}} 
\, {z_u}\, ({\slash P } - M_d)^{-1}. \nonumber  
\end{eqnarray}
Note that since left handed up ($I_3 = 1/2$) and down ($I_3 = -1/2$)
components belong to the same multiplet and $Z_L$ has already been fixed
by Eq. (\ref{yy}) we cannot have $z_u = 1$.
The residue  $z_u$  is finite and is given by
\begin{eqnarray}
z_u \, \equiv \, 1 + {\delta z_u} \,=\,
 1\,+\, ({{\alpha_s} \over {2 \pi}})  \, c_F \,
 ({{\Pi}_u}  (M^2_u)- {{\Pi}_d}  (M^2_d)).
\end{eqnarray}
The  $ {{\Pi}_u}  (M^2_u) $ appearing above is in form identical to
$ {{\Pi}_d}  (M^2_d) $  defined in Eq. (\ref{aa}) with the replacements
$M_d ,m_{{\tilde d}_L} ,m_{{\tilde d}_L^c} \rightarrow
  M_u ,m_{{\tilde u}_L} ,m_{{\tilde u}_L^c}$.

Since  $ {\delta z_u} \neq 1 $ we have an additional contribution to the
amplitude which stems from the wave function renormalization of the external
up quark line; at the one-loop level this is given by
\begin{eqnarray}
{{\cal A}_2} &=& i\;{g \over \sqrt{2}} M_0 \; 
		 ({ {\delta z_u} \over 2} ) \nonumber \\  
      &=& i\; c_F \, {  {g {\alpha_s} }   \over {4 \pi \sqrt{2}   }  }  
 ({{\Pi}_u}  (M^2_u)- {{\Pi}_d}  (M^2_d))   \, M_0. 
\end{eqnarray}
This completes our calculation of the SQCD corrections to the amplitude 
for $W^+ \to u {\bar d} $.

We now proceed to discussing the corrections to the hadronic width of the 
$W$-boson. The one-loop hadronic width $\Gamma ^{(1)}$ can be written
as 
\begin{eqnarray}
\Gamma ^{(1)} \,=\,{\Gamma_{u{\bar d}} ^{(0)}} \, (1\,+\,{\delta }),
\end{eqnarray}
where   ${\Gamma_{u{\bar d}} ^{(0)}} $  
is the tree level hadronic width for one family. In the limit of vanishing
quark masses this is given by  
${\Gamma_{u\bar d} ^{(0)}} \,=\,{ {\alpha}_w } M_W / 4$. 
The SQCD corrections to ${\delta }$ can be found from the 
amplitudes ${\cal A}_1, {{\cal A}_1}',{\cal A}_2$ we have just calculated.
It is found that
\begin{eqnarray}
\delta^{SQCD} \, = \, {{\alpha_s} \over { \pi}}  \, c_F \, 
[\,2 C_{24} + B_1 +{{\Pi_u -\Pi_d} \over 2}\,] + ... 
\end{eqnarray}
In order to avoid confusion we should say that $\delta^{SQCD}$ accounts
for only the supersymmetric corrections, that is those due to the exchange
of gluinos and squarks.
The functions $C_{24}, B_1, \Pi_{u,d}$ are as they appear in the definitions
of the amplitudes ${\cal A}_{1},{\cal A}_{1}',{\cal A}_{2}$, while the ellipses
denote terms proportional to the external quark masses. In the limit
of vanishing quark masses{\footnote{The ${\cal O} (M_{u,d}) $ 
terms give a negligible contribution
and hence it is permissible to omit them. }},
$\delta^{SQCD}$ can be cast in the following integral form 
\begin{eqnarray}
{\delta }^{SQCD} \,&=&\,{{ 2 \alpha_s} \over {3 \pi}}
{\int_{0}^{1}xdx \int_{0}^{1}dy}  \nonumber \\
&&\times \, {\ln \,\{ 
{  {(x m^2_{{\tilde u}_L}+(1-x) M^2_{{\tilde g}}) \,
(x m^2_{{\tilde d}_L}+(1-x) M^2_{{\tilde g}})} \over
{(M^2_W x^2 y(y-1)+(m^2_{{\tilde d}_L}-m^2_{{\tilde u}_L})xy+
(m^2_{{\tilde u}_L}- M^2_{{\tilde g}})x+M^2_{{\tilde g}})^2}    }}\}.
\end{eqnarray}
>From the form above we can easily get first estimates of the magnitude
of the SQCD corrections as will be seen in the sequel.

\begin{figure}[t]
\setlength{\unitlength}{1cm}
\begin{center}
\epsfxsize=22.cm
\epsffile[70 300 700 500]{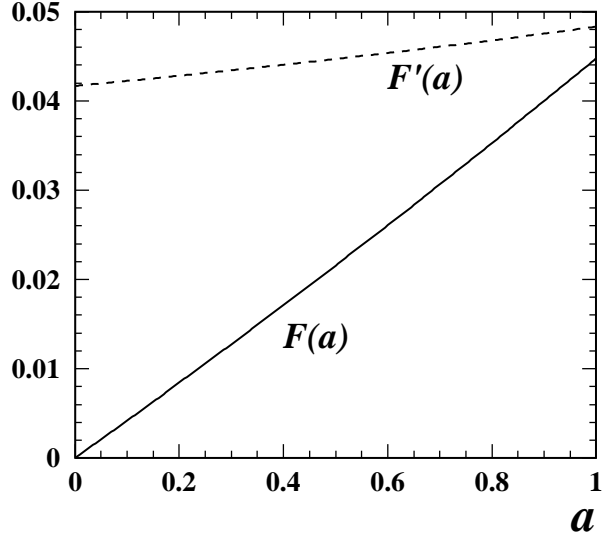}
\begin{minipage}[t]{14.7cm}
\caption[] {The function $F(a)$ and its derivative as described
in the text.} 
\end{minipage}
\end{center}
\end{figure}
In order to simplify the discussion let us assume that
$ m_{{\tilde u}_L} \approx m_{{\tilde d}_L} \approx M_{{\tilde g}} = M_S$,
where $M_S$ sets the order of the supersymmetry breaking scale. In
this case the expression above for ${\delta}^{SQCD}$ is simplified, to become
\begin{eqnarray}
{\delta }^{SQCD} \,=\,{{ 4 \alpha_s} \over {3 \pi}}
{\int_{0}^{1}xdx \int_{0}^{1}dy} 
  \ln \, 
 \{ {{M_S^2} \over { M^2_W x^2y(y-1)+ M_S^2}}  \}  \nonumber \\ \equiv
{{ 4 \alpha_s} \over {3 \pi}} F(M^2_W / M_S^2).
\end{eqnarray}
Since $M_S>M_W$ the function   $F(M^2_W / M_S^2)$
can be expanded in powers of 
$a \equiv  M^2_W / M_S^2$. The result of such an expansion is 
\begin{eqnarray}
F(a)={a \over 24}+{{a^2} \over 360} +{{a^3} \over 3360}+ ...
\end{eqnarray}
Keeping the first term of the expansion results in
\begin{eqnarray}
{\delta }^{SQCD} \,&=&\,{{ \alpha_s} \over {\pi}}
{1 \over 18} { {M^2_W} \over {M_S^2}}.
\end{eqnarray}
${\delta }^{SQCD}$ is very well approximated by keeping only
the leading term in the expansion of $F$ as can be seen from Fig. 2
where both the function and its derivative are plotted. In fact
the function $F(a)$ is almost linear in the interval $0<a<1$ 
with almost constant derivative, justifying the linear approximation
to $F$ which led to the result above. Note the appearance of an extra
suppression factor $1/18$ in addition to the expected
${ {M^2_W} \over {M_S^2}}$ factor
due to the decoupling of sparticles as we pass below the scale $M_S$.
This situation persists also in other cases as for instance when
the gluino is lighter than the squarks, i.e
$ M_{\tilde g} \ll  m_{{\tilde u}_L,{\tilde d}_L } $. In that case
employing the fact that 
$\Delta m^2 \equiv m^2_{{\tilde d}_L} - m^2_{{\tilde u}_L} \ll
    m^2_{{{\tilde u}_L},{{\tilde d}_L} } \equiv M_S^2 $, 
since the difference of the masses squared of the 
${\tilde u}_L, {\tilde d}_L$ squarks 
is of the order of the electroweak scale{\footnote {We assume universal
boundary conditions for the squarks at the unification scale.}, we get in
an analogous way exactly the same result with the suppression factor
$1/18$ being replaced by $2/27$. 
Our complete numerical analysis uses the full expression for
$\delta^{SQCD}$ and has
actually covered the whole parameter space of the MSSM assuming
universal boundary conditions for the squark masses at the unification 
point where the couplings merge. In all cases the SQCD corrections turned 
out to be of the order 
${\cal O} (5\%) {{ \alpha_s} \over {\pi}}  { {M^2_W} \over {M_S^2}} $ 
or less, instead of the expected
${{ \alpha_s} \over {\pi}} { {M^2_W} \over {M_S^2}} $ behaviour. 
If we compare this with ${{ \alpha_s} \over {\pi}}$, 
which is the contribution of gluons to $\delta $, we see that 
the appearance of gluinos and squarks has a negligible effect 
$ \leq {\cal O}(10^{-2}){{ \alpha_s} \over {\pi}} $ on the hadronic width
of the $W$-boson. Actually in the
constrained MSSM with radiative symmetry breaking, these corrections turn 
out to be even smaller 
$ {\cal O}(10^{-3} - 10^{-4}){{ \alpha_s} \over {\pi}} $.
Therefore 
supersymmetric QCD corrections to the $W$-boson width are at best of the 
order of the two-loop electroweak corrections and not of relevance
to current experiments.

\vspace{5.5cm}
\noindent
{\large {\bf Acknowledgements}}  
\\ 
We  want to  thank  James  Stirling  for a  careful  reading  of  this
manuscript. This work was supported by the EU Human  
Capital and Mobility  
Programme, CHRX--CT93--0319.  
 
\newpage


\begin{thebibliography}{99}

\bibitem{cern}`Determination of the Mass of
the $W$ Boson', Z.~Kunszt and W.J.~Stirling et al.,
in {\it Physics at LEP2}, eds. G.~Altarelli, T.~Sj\"ostrand  and F.~Zwirner,
CERN Report 96-01 (1996), vol.1, p.~141;\\
`WW cross-sections and distributions', 
W. Beenakker and F.A. Berends et al.,
in {\it Physics at LEP2},
eds. G.~Altarelli, T.~Sj\"ostrand  and F.~Zwirner,
CERN Report 96-01 (1996), vol. 1, p.~81.


\bibitem{chang} T.H.~Chang, K.J.F.~Gaemers and W.L.~van~Neerven,
 Nucl. Phys. {\bf B202} (1982) 407;\\
 K.~Inoue, A.~Kakuto, H.~Komatsu and S.~Takeshita, Prog. Theor. Phys. 
 {\bf 64} (1980) 1008;\\
 D.~Bardin, S.~Riemann and T.~Riemann, Z. Phys. C {\bf 32} (1986) 121;\\
 J.W.~Jun and C.~Jue, Mod. Phys. Lett. {\bf A6} (1991) 2767.
 
 
\bibitem{denner} A.~Denner and T.~Sack, Z. Phys. C {\bf 46} (1990) 653.

\bibitem{shin} D.-S.~Shin, Nucl. Phys. {\bf B449} (1995) 69. 

\bibitem{rosner} J.~Rosner, M.~Worah and T. Takeuchi,
 Phys. Rev. {\bf D49} (1994) 1363. 

\bibitem{on-shell} D.A.~Ross and J.C.~Taylor, Nucl. Phys. {\bf B51} (1973) 
 125.

\bibitem{passa} G.~Passarino and M.~Veltman, 
Nucl. Phys. {\bf B160} (1979) 151. 

\bibitem{hollik} M.~B\"{o}hm, H.~Spiesberger and W.~Hollik, Fortschr.
Phys. {\bf 34} (1986) 687.

\bibitem{hollik2} ``Renormalization of the Standard Model",
W.~Hollik, appears in {\it Precision tests of the Standard Model},
 Advanced series on directions in high-energy physics, ed. Paul Langacker,
World Scientific, 1993. 

\bibitem{hooft} G.~'t Hooft and M.~Veltman, Nucl. Phys. {\bf B153} (1979)
369.


\end{thebibliography}
\end{document}